\documentclass{ifacconf}

\usepackage{graphicx}      
\usepackage{natbib}        

\usepackage{amsmath,amssymb,amsfonts}
\usepackage{mathrsfs}
\usepackage{cases}
\usepackage{array}
\usepackage{graphicx}
\usepackage{pgfplots}
\usepackage{tikz}
\pgfplotsset{compat=1.9}
\usepgfplotslibrary{groupplots}
\usepackage{algorithm,algorithmic}

\usepackage{enumitem} 
\usepackage{bm}
\usepackage{makecell}

\usepackage[draft]{todonotes}

\newcommand{\change}[1]{\textcolor{black}{#1}}

\newtheorem{remark}{Remark}
\newtheorem{assumption}{Assumption}

\DeclareMathOperator{\R}{\mathbb{R}}
\DeclareMathOperator{\N}{\mathbb{N}}

\usepackage{url}

\usepackage{eso-pic}
\AddToShipoutPictureFG*{%
  \AtPageUpperLeft{%
    \put(48,-60){%
      \parbox{0.83\paperwidth}{%
        \footnotesize
        \textbf{Extended version of accepted paper} for IFAC World Congress 2026.\par
        \copyright\ 2026 the authors. This work has been accepted to IFAC for publication under a Creative Commons Licence CC-BY-NC-ND.
      }%
    }%
  }%
}

\begin{document}
\begin{frontmatter}

\title{Disturbance-Adaptive Model Predictive Control for Bounded Average Constraint Violations\thanksref{footnoteinfo}}

\thanks[footnoteinfo]{This work was supported by the Swiss National Science Foundation (SNSF) under the NCCR Automation project, grant agreement 51NF40\_180545. \textbf{This is the extended version of the accepted IFAC paper.}}

\author[First]{Jicheng Shi} 
\author[First]{Colin N. Jones} 

\address[First]{Automatic Laboratory, EPFL, 1015 Lausanne, Switzerland (e-mail: $\{$jicheng.shi, colin.jones$\}$@epfl.ch).}

\begin{abstract}                
This paper considers stochastic linear time-invariant systems subject to time-averaged state-constraint violation bounds, without assuming knowledge of the disturbance distribution. We propose a disturbance-adaptive model predictive control (DAD-MPC) framework, which adjusts the confidence level and the induced disturbance set based on measured constraint violations. Using a robust invariance method, DAD-MPC ensures recursive feasibility and guarantees robust or asymptotic bounds on the average violation rate. Additionally, the bounds remain valid even with an inaccurate disturbance model, enabling the use of data-driven disturbance quantification methods such as conformal prediction. Simulation results demonstrate that the proposed approach reduces closed-loop cumulative cost compared to state-of-the-art methods across different target violation rates, while satisfying average violation bounds.
\end{abstract}


\end{frontmatter}

\section{Introduction}
Model Predictive Control (MPC) is a popular control strategy due to its natural integration of control objectives and constraints~\citep{morari1999model}. MPC uses a model to predict and optimize system performance over a future horizon. Operating in a receding-horizon manner, it enhances the robustness and efficiency of systems and is widely applied across industries~\citep{qin2003survey}.
Despite its numerous advantages, uncertainties in prediction, such as model mismatch and external disturbances, present a significant challenge, which can substantially impact the optimality and reliability of MPC~\citep{morari1999model}.

One approach to addressing these uncertainties is through robust MPC~\citep{bemporad2007robust}, which leverages robust optimization methods and uncertainty bounds to ensure strict constraint satisfaction. Numerous theoretical studies have explored various robust MPC schemes with guarantees~\citep{mayne2006robust,goulart2006optimization}. Despite its theoretical guarantees, robust formulations can lead to excessive conservatism, affecting both the control performance and the region of feasibility~\citep{qin2003survey}. Consequently, soft constraints are often employed in practical applications of robust MPC methods as a compromise between control performance and conservatism~\citep{schwenzer2021review,zeilinger2014soft}. 

To mitigate the conservatism observed in robust MPC, stochastic MPC offers a viable alternative by allowing occasional constraint violations managed through chance constraints with specified probabilities~\citep{mesbah2016stochastic,lorenzen2016constraint,MPCsto_korda2011strongly}. For instance, the probability of constraint violation at each time step can be constrained as described in~\citet{MPCsto_korda2011strongly}. This approach can result in improved control costs compared to robust MPC.  However, the least-restrictive formulation proposed in~\citet{MPCsto_korda2011strongly} can exhibit conservatism under some conditions, as highlighted in~\citet{korda2012stochastic}. To further reduce conservatism, an alternative is to impose a bound on the time-averaged state-constraint violations. This is practically as expressive as chance constraints~\citep{korda2012stochastic} and is commonly used in applications, such as average comfort violations in building climate control~\citep{shi2026disturbance,lian2023adaptive}, and fatigue constraints in wind turbine control~\citep{cannon2009probabilistic}.


Despite its potential practicality, the time-averaged violation bound has not been extensively researched with limited theoretical studies~\citep{korda2012stochastic,oldewurtel2013adaptively,fleming2017time}. In~\citet{korda2012stochastic}, the authors modify inequality constraints based on the current average number of violations at each sampling step. The resulting method guarantees a probability of average constraint violation using a stochastic invariance method, which requires knowledge of the disturbance distribution. \citet{oldewurtel2013adaptively} propose an online adaptation of constraint tightening based on the observed average constraint violation. It establishes an asymptotic bound on the average violation for a specific class of linear systems with one-step controllability. Similarly, \citet{fleming2017time} designs an inequality constraint adaptation method for linear systems with parameter and external uncertainties. It guarantees a robust average violation bound over any time period under affine state feedback and a specific type of state constraints.



\begin{figure}[!t]
    \centering
    \includegraphics[width=1.0\linewidth]{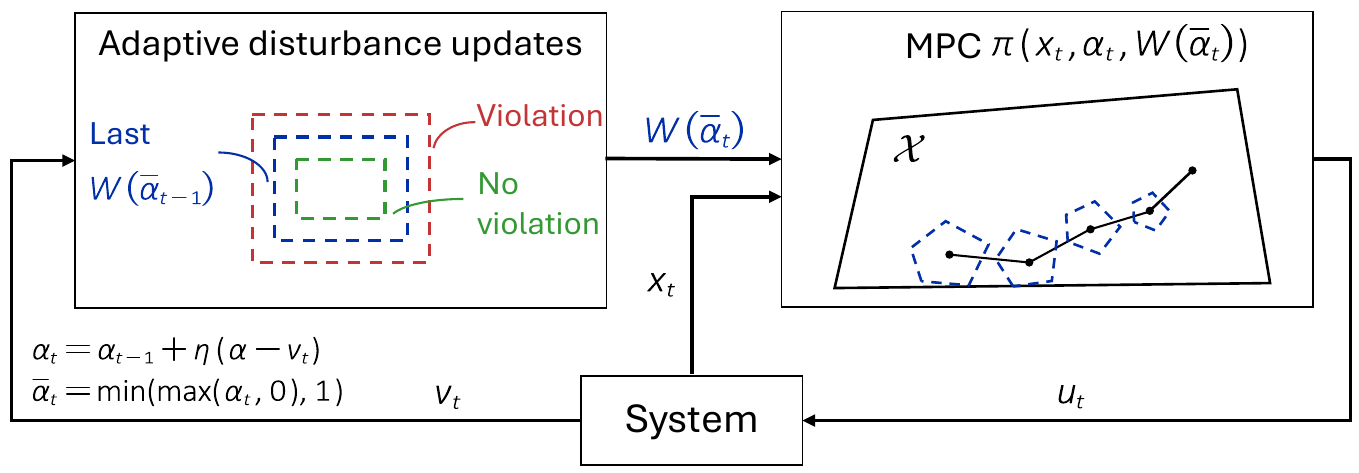}
    \vspace{-18pt} 
    \caption{\footnotesize Illustration of DAD-MPC: Based on the violation indicator $v_t$, the variable $\alpha_t$ and the clipped confidence variable $\bar{\alpha}_t \in [0,1]$ are updated, which  determines the estimated disturbance set $W(\bar{\alpha}_t)\subseteq\mathbb{W}$. \change{Then the MPC policy $\pi(x_t,\alpha_t,W(\bar{\alpha}_t))$ computes the input $u_t$, which is applied to the real system.}}
    \label{fig:dad_concept}
\end{figure}

In this paper, we propose a DAD-MPC framework for enforcing time-averaged violation bounds under weaker assumptions than existing methods. It enables a flexible MPC design without requiring knowledge of the disturbance distribution. As illustrated in Figure~\ref{fig:dad_concept}, \change{the main idea is to adapt a confidence variable and the induced disturbance set within the MPC based on the current violation indicator.} A follow-up work~\citep{shi2026disturbance} applies this framework to data-driven building climate control, which reduces energy savings under user-specified comfort violation bounds. The contributions of this work are summarized in four points.
\begin{itemize}[nolistsep,leftmargin=1.0em]
\item The proposed DAD-MPC applies to stochastic linear time-invariant (LTI) systems with flexible policy choices.
\item It ensures recursive feasibility and guarantees asymptotic or robust bounds on average violation rate using a robust invariance approach.
\item It does not require exact knowledge of the disturbance distribution. Instead, we quantify disturbances using data-driven conformal prediction~\citep{cp_vovk2005}.
\item Its efficacy is validated through simulations, outperforming state-of-the-art methods.
\end{itemize}

\noindent\textbf{Notation}:
The $i$-th element in a vector $z$ is denoted by $z(i)$. 
$\N$ and $\N_{+}$ represent the sets of non-negative and positive integers, respectively. $\N_i^j$ denotes the set of consecutive integers $\{i, i + 1, \dots, j\}$. $\mathbf{0}$ denotes a zero vector/matrix with a proper size.

\section{Problem Statement}
We consider a dynamical system described by the following LTI model:
\begin{equation}
\begin{aligned} \label{eqn:LTI_sys}
    x_{t+1} = Ax_t + Bu_t + w_t
\end{aligned}    
\end{equation}
where the state $x_t \in \R^{n_x}$, control input $u_t \in \R^{n_u}$ and the disturbance $w_t \in \R^{n_x}$. Note that $w_t$ is not necessarily independent and identically distributed (i.i.d.) in this work.

\begin{assumption} \label{assump:w_bound}
The state $x_t$ is measured at each sampling time $t$. The system~\eqref{eqn:LTI_sys} is stabilizable.
\change{The disturbance $w_t$ belongs to a known compact polyhedral disturbance set $\mathbb{W}\subset\mathbb{R}^{n_x}$, for all $t\in\mathbb{N}$ (not necessarily the tight support of $w_t$).}
\end{assumption}




The system~\eqref{eqn:LTI_sys} is required to satisfy constraints on the inputs and states. The input constraint is defined as
\begin{align*}
    u_t \in \mathcal{U}, \;t\in \N_{}.
\end{align*}
The state $x_t$ should remain within a polyhedron: 
\begin{align} \label{eqn:cons_x}
    \mathcal{X} =\{x | F_x x \leq f_x \}.
\end{align} but occasional violations are allowed. We define a binary variable $v_t$ to indicate if the current state constraint is violated:
\begin{align*}
    v_t = \begin{cases}
        0, & \text{if } x_t \in \mathcal{X} \\
        1, & \text{if } x_t \notin \mathcal{X}\enspace.
    \end{cases}
\end{align*}
We consider two types of average violation bounds on the state $x_t$. The asymptotic type is:
\begin{align} \label{eqn:cons_vio_asy}
    \lim_{t\rightarrow\infty} \frac{\sum_{i=1}^{t} v_i}{t} \leq  \alpha,
\end{align}
and the robust type is:
\begin{align} \label{eqn:cons_vio_rob}
    \frac{\sum_{i=1}^{t} v_i}{t} \leq  \alpha, \;\forall \; t \in \N_{+}.
\end{align}
For both constraints, $\alpha \in [0, 1)$ is a user-defined parameter that bounds the averaged violation rate. \change{For example, in building climate control, 5\% average time outside the comfort constraints is permitted (European standard EN15251 Annex G, \citet{comite2007indoor}).} The robust type~\eqref{eqn:cons_vio_rob} is stricter, requiring bound satisfaction at all times.

Additionally, we introduce other types of violation constraints for comparison. The chance constraint type, often used in stochastic MPC, is:
\begin{align} \label{eqn:cons_vio_stompc}
    Pr(v_t =1) = Pr(x_t \notin \mathcal{X} ) \leq \alpha, \; \forall \;t\in \N_{+}.
\end{align}
These constraints are often enforced via sufficient conditional counterparts~\citep{mesbah2016stochastic,MPCsto_korda2011strongly}.
For example, \citet{MPCsto_korda2011strongly} uses:
\begin{align*} 
    Pr(v_{t+1} =1|x_t) \leq \alpha, \; \forall \;t\in \N_{},
\end{align*} conditioned on the last state. However, this can lead to more conservative control performance~\citep{korda2012stochastic}.  
\citet{korda2012stochastic} propose a less conservative constraint by focusing on the average probability of violation:
\begin{align} \label{eqn:cons_vio_milan}
    \frac{\sum_{i=1}^{t} Pr(v_i =1)}{t}  \leq \alpha, \; \forall \;t\in \N_{+}.
\end{align}
The asymptotic condition~\eqref{eqn:cons_vio_asy} focused on in this work is also less conservative, since, under mild regularity (e.g., stationarity/ergodicity of ${v_t}$), \eqref{eqn:cons_vio_stompc} implies~\eqref{eqn:cons_vio_asy} almost surely.


\section{Disturbance-adaptive Model Predictive Control}

This section introduces the DAD-MPC framework to satisfy the average violation bounds~\eqref{eqn:cons_vio_asy} and~\eqref{eqn:cons_vio_rob}. 
\change{Section~\ref{sect:how_dad_work} describes the online operation of DAD-MPC (Algorithm~\ref{alg:dad_mpc}). As illustrated in Fig.~\ref{fig:dad_concept}, the framework expands the estimated disturbance set when a violation occurs and shrinks it otherwise.} 

Section~\ref{sect:guarantee} develops a first-step robust invariance (FRI) approach, inspired by~\citet{korda2012stochastic}, to establish guarantees for~\eqref{eqn:cons_vio_asy} and~\eqref{eqn:cons_vio_rob}. The core idea is that if the average violation rate up to time $t$ is too large, the controller switches to the largest disturbance set $\mathbb{W}$ and enforces $x_{t+1}\in\mathcal{X}$ for all $w_t\in\mathbb{W}$.

\change{The FRI approach allows a flexible design for the control policy. Section~\ref{sect:DAD-MPC} presents a design guidance on the control policy, which establishes a violation feedback loop. As an example, we present a concrete formulation of the DAD-MPC control policy using conformal prediction~\citep{cp_vovk2005} and disturbance-affine MPC~\citep{goulart2006optimization}.}



\subsection{DAD-MPC framework} \label{sect:how_dad_work}

DAD-MPC consists of three key components
\begin{itemize}[nolistsep,leftmargin=2.0em]
    \item A time-varying variable $\alpha_t \in \mathbb{R}$ that adapts according to the violations; \change{A clipped confidence variable $\bar{\alpha}_t = \min(\max(\alpha_t,0),1)$ $\in [0,1]$;}
    \item \change{A disturbance-set estimator $W(\cdot)$ that maps $\bar{\alpha}_t$ to a bounded disturbance set $W(\bar{\alpha}_t)$;}
    \item An MPC control policy denoted $\pi(x_t,\alpha_t, W(\bar{\alpha}_t))$, which depends on the system state $x_t$, the variable $\alpha_t$ and the estimated set $W(\bar{\alpha}_t)$. \change{For notational simplicity, we denote it as $\pi(\cdot)$ throughout the paper.}
\end{itemize}
The online operation of DAD-MPC is outlined in Algorithm~\ref{alg:dad_mpc}. It adaptively updates the variable $\alpha_t$ in~\eqref{eqn:dad_alpha}, dynamically adjusting the estimated disturbance set in $\pi(\cdot)$. 
This adaptation distinguishes DAD-MPC from traditional robust and stochastic MPC methods with fixed constraint-tightening values~\citep{bemporad2007robust,mesbah2016stochastic}. \change{In the next section, we build $\pi(\cdot)$ via the FRI approach and show that the violation bounds~\eqref{eqn:cons_vio_asy} and~\eqref{eqn:cons_vio_rob} hold for any estimator $W(\cdot)$.}



\begin{algorithm} 
  \caption{DAD-MPC framework} \label{alg:dad_mpc}
{
\textbf{Input:} Target average violation $\alpha$, initial confidence value $\alpha_0$, constant update rate $\eta > 0$,  disturbance-set estimator $W(\cdot)$, control policy $\pi(\cdot)$
\begin{itemize} [nolistsep,leftmargin=1.5em]
    \item[1)] At $t>0$, update $\alpha_t$ based on the current violation indicator $v_t$: 
    {\setlength{\abovedisplayskip}{1.6pt}\setlength{\belowdisplayskip}{1.6pt}
    \begin{align} \label{eqn:dad_alpha}
        \alpha_{t} = \alpha_{t-1} + \eta(\alpha - v_t).
    \end{align}
    }    
    \item[2)] \change{Compute the clipped confidence variable $\bar{\alpha}_t = \min(\max(\alpha_t,0),1)$. Determine the current estimation of the disturbance set by $W(\bar{\alpha}_t)$.}    
    \item[3)] Retrieve the current state measurement $x_t$ and apply the MPC control policy by $u_t = \pi(\cdot)$. 
    \item[4)] Pause until the next sampling time, update $t\leftarrow t+1$ and return to step 1).    
\end{itemize}
}
\end{algorithm}


\subsection{FRI-based control policy and  violation bound guarantees}  \label{sect:guarantee}

\change{We now specify $\pi(\cdot)$ via the FRI approach. By Theorem~\ref{thm:FRI}, this guarantees the bounds~\eqref{eqn:cons_vio_asy} and~\eqref{eqn:cons_vio_rob}.}

\subsubsection{FRI-based control policy} \label{sect:FRI}

Define the pre-set operator of a set $\mathcal{M} \subseteq \R^{n_x}$ w.r.t.\ the control model~\eqref{eqn:LTI_sys} and $\mathcal{U}$ as:
\begin{align*}
    & \text{Pre}(\mathcal{M}) \\
    &= \left\{ x \in \mathbb{R}^{n_x} \;\middle|\;
        \exists u \in \mathcal{U} \text{ s.t. } Ax + Bu + w \in \mathcal{M},\; \forall w \in \mathbb{W}
    \right\}.
\end{align*}
One key component of the FRI method is the pre-set of the state constraint set, $\mathcal{X}_r:=\text{Pre}(\mathcal{X})$.
\change{The second is a robust controlled invariant (RCI) subset $\mathcal{S}_1\subseteq\mathcal{X}_r$ satisfying:}
\begin{equation} \label{eqn:rci}
\begin{aligned}
    \forall x \in \mathcal{S}_1, \exists u \in \mathcal{U} \; \text{s.t. } &Ax + Bu + w \in \mathcal{S}_1 \cap \mathcal{X}, \ \forall w \in \mathbb{W}
\end{aligned}
\end{equation}
In addition, consider a sequence of pre-sets of length $n_s$ starting from $\mathcal{S}_1$:
\begin{align*}
    \mathcal{S}_{k+1} := \text{Pre}(\mathcal{S}_k), \; k = \N_1^{n_s-1} .
\end{align*}
Please note that $\mathcal{S}_1$ is not the standard RCI for $\mathcal{X}$ and is not necessarily a subset of the state constraint $\mathcal{X}$. We assume the existence of $\mathcal{S}_1$.
\begin{assumption} \label{assump:rci}
    \change{For the chosen $\mathbb{W}$,} a nonempty RCI subset $\mathcal{S}_1 \subseteq \mathcal{X}_r$ exists and has been characterized.
\end{assumption} 
 
Based on these elements, define the FRI-based admissible input set:
\begingroup
\allowdisplaybreaks
\begin{subequations} \label{eqn:FRI_u_cons}
    \begin{align}
        \mathbb{U}&(x_t, \alpha_t) = \left\{ u \in \mathcal{U} \; \text{s.t.} \right. \nonumber \\
        & r_t:=\max\left\{ \min\left\{ \left\lfloor \frac{\alpha_t - \alpha_{\text{low}}}{\eta(1-\alpha)} \right\rfloor, n_s \right\}, 1 \right\}, \label{eqn:r_t} \\
        & Ax_t + Bu_t + w \in \mathcal{S}_{r_t}, \; \forall w \in \mathbb{W} \label{eqn:FRI_u_cons_1} \\
        & \text{if} \; 
        \begin{aligned}[t]
        & \alpha_t < \alpha_{\text{low}} + \eta(1-\alpha) \Rightarrow \\
        & Ax_t + Bu_t + w \in \mathcal{X},
        \forall w \in \mathbb{W}\}.
    \end{aligned}
        \label{eqn:FRI_u_cons_2}
    \end{align}
\end{subequations}
\endgroup
Here $\alpha_{\text{low}}$ is user-defined.
\change{The MPC control policy $\pi(\cdot)$ is then formulated in a flexible way:}
\begin{equation} \label{eqn:dadmpc_policy}
\begin{aligned}
    \min \; & J(\cdot) \\
    \text{s.t.} \; & u_{0|t} \in \mathbb{U}(x_t, \alpha_t)
\end{aligned}
\end{equation}
where $J(\cdot)$ is a generic cost function.

\subsubsection{Violation bound guarantees}
\change{We now establish satisfaction of~\eqref{eqn:cons_vio_asy} and~\eqref{eqn:cons_vio_rob} under a suitable choice of $\alpha_0$ in Theorem~\ref{thm:FRI}.}
Define the feasibility set of $\mathbb{U}(x_t, \alpha_t)$ as:
\begin{align*}
    \Pi  &= \left\{ (x_t, \alpha_t) \; \middle| \; \mathbb{U}(x_t,\alpha_t) \neq \emptyset
    \right\}.
\end{align*}

\begin{thm} \label{thm:FRI}
    Consider system~\eqref{eqn:LTI_sys} controlled by Algorithm~\ref{alg:dad_mpc} with  user-defined finite parameters $\alpha_0$ and $\eta$.
    If Assumptions~\ref{assump:w_bound}, \ref{assump:rci} hold and the control action satisfies $u_t=\pi(\cdot) \in \mathbb{U}(x_t, \alpha_t)$, the following holds:
    \begin{itemize} [nolistsep,leftmargin=2.5em]
        \item[I.] For a chosen $\alpha_0$ and the corresponding $r_0$ from~\eqref{eqn:r_t}, if $x_0\in \mathcal{S}_{r_0}$, then $(x_0, \alpha_0) \in \Pi$;
        \item[II.] If $(x_t, \alpha_t) \in \Pi$, then $(x_{t+1}, \alpha_{t+1}) \in \Pi$, for $t\in \N_+$;
        \item[III.] If $(x_0, \alpha_0) \in \Pi$ and $\alpha_{\text{low}}$ is finite, then $\lim_{t\rightarrow\infty}\frac{\alpha_t}{t} \geq 0$ and~\eqref{eqn:cons_vio_asy}  holds; If additionally $\alpha_0 \leq \alpha_{\text{low}}$, then $\alpha_t \geq \alpha_0$, $ \forall t \in \N_{+}$ and~\eqref{eqn:cons_vio_rob} holds.
    \end{itemize}    
\end{thm}

\change{The proof of Theorem~\ref{thm:FRI}.III is based on the following Lemmas~\ref{lem:dad_bound} and~\ref{lem:bound}. Firstly, we establish that the average violation $\frac{\sum_{i=1}^{t} v_i}{t}$ is linearly related to $\frac{\alpha_t}{t}$ by Lemma~\ref{lem:dad_bound}.}
\begin{lem} \label{lem:dad_bound}
    Consider system~\eqref{eqn:LTI_sys} controlled by Algorithm~\ref{alg:dad_mpc}. Then, at time $t$, the average violation satisfies:
    \begin{align} \label{eqn:thm_relation}
        \frac{\sum_{i=1}^{t} v_i}{t} = \alpha + \frac{\alpha_0-\alpha_t}{t\eta}
    \end{align} 
    As a result, the average violation is bounded by:
    \begin{align} \label{eqn:thm_bound}
    \frac{\sum_{i=1}^{t} v_i}{t} \in \alpha + \left[ \frac{\alpha_0-\alpha_{max,t}}{t\eta},\; \frac{\alpha_0-\alpha_{min,t}}{t\eta} \right],
    \end{align}
    where $\alpha_{max,t}:= \max_{i\in\N_0^t}\alpha_i, \; 
    \alpha_{min,t} := \min_{i\in\N_0^t}\alpha_i$.
\end{lem}

\begin{pf}
    The update equation~\eqref{eqn:dad_alpha} can be expanded recursively from time $0$ to $t$:
        $\alpha_{t} = \alpha_{0} + \eta\sum_{i=1}^{t}(\alpha - v_i)$. Rearranging this equation yields~\eqref{eqn:thm_relation}. Given that $\alpha_t \in \left[ \alpha_{min,t},\; \alpha_{max,t} \right]$, we obtain the bound in~\eqref{eqn:thm_bound}.  
\end{pf}

\change{Lemma~\ref{lem:bound} then provides equivalent conditions for~\eqref{eqn:cons_vio_asy} and~\eqref{eqn:cons_vio_rob}.}

\begin{lem}
\label{lem:bound}
Consider system~\eqref{eqn:LTI_sys} controlled by Algorithm~\ref{alg:dad_mpc} with  user-defined finite parameters $\alpha_0$ and $\eta$. The following two statements are equivalent:
\begin{itemize}[nolistsep,leftmargin=1.5em]
    \item The asymptotic violation bound~\eqref{eqn:cons_vio_asy} is satisfied;
    \item \textbf{C1}: $\lim_{t\rightarrow\infty}\frac{\alpha_t}{t} \geq 0$.
\end{itemize}
And the following two statements are equivalent:
\begin{itemize}[nolistsep,leftmargin=1.5em]
    \item The robust violation bound~\eqref{eqn:cons_vio_rob} is satisfied;
    \item \textbf{C2}: $\alpha_t \geq \alpha_0$, $ \forall t \in \N_{+}$.
\end{itemize}
\end{lem}
\begin{pf}
Taking the limit of the relation~\eqref{eqn:thm_relation} from Lemma~\ref{lem:dad_bound} gives:
    \begin{align*} \lim_{t\rightarrow\infty}\frac{\sum_{i=1}^{t} v_i}{t} &= \lim_{t\rightarrow\infty} \alpha + \frac{\alpha_0-\alpha_t}{t\eta} = \alpha - \lim_{t\rightarrow\infty}\frac{\alpha_t}{t\eta},
    \end{align*}
    because $\alpha_0$ and $\eta$ are finite. Therefore, the asymptotic bound~\eqref{eqn:cons_vio_asy} holds if and only if $\lim_{t \to \infty} \frac{\alpha_t}{t} \geq 0$, which is \textbf{C1}.
    For the robust bound, again using the relation~\eqref{eqn:thm_relation}, it is obvious to see that~\eqref{eqn:cons_vio_rob} and \textbf{C2} are equivalent.
\end{pf}

\change{We now prove Theorem~\ref{thm:FRI}.}
\begin{pf}[for Theorem~\ref{thm:FRI}]
By invariance preservation of the pre-set operator, $\mathcal{S}_k$ is RCI for all $k\in\{1,\dots,n_s\}$. By construction, $\mathcal{S}_k \subseteq \operatorname{Pre}(\mathcal{S}_k)=\mathcal{S}_{k+1}$, so the sequence is nested.\\
    I. At time $0$, if $x_0\in \mathcal{S}_{r_0}$, then~\eqref{eqn:FRI_u_cons_1} is feasible because $\mathcal{S}_{r_0}$ is an RCI set. \change{If \eqref{eqn:FRI_u_cons_2} is active,  then by~\eqref{eqn:r_t} we have $r_0=1$, hence $x_0\in\mathcal{S}_1$. By~\eqref{eqn:rci},  \eqref{eqn:FRI_u_cons_2} is feasible.}
    Thus, $(x_0, \alpha_0) \in \Pi$. \\
    II. If $(x_t,\alpha_t)\in\Pi$, then by~\eqref{eqn:FRI_u_cons_1} we have $x_{t+1}\in\mathcal{S}_{r_t}$. For~\eqref{eqn:FRI_u_cons_1} at time $t+1$:\\
    - If $r_{t+1}=r_t-1$, feasibility holds because $\mathcal{S}_{k+1} := \text{Pre}(\mathcal{S}_k), \forall k = \N_1^{n_s}$. \\
    - If $r_{t+1}=r_t$, feasibility holds since $\mathcal{S}_{k}$ is an RCI set $\forall k = \N_1^{n_s}$. \\
    - If $r_{t+1}>r_t$, feasibility is preserved due to the nested property $\mathcal{S}_{r_t} \subseteq \mathcal{S}_{r_{t+1}}$. \\
    \change{If~\eqref{eqn:FRI_u_cons_2} is activated at time $t+1$, i.e., $\alpha_{t+1} < \alpha_{\text{low}} + \eta(1-\alpha)$,  then by the update $\alpha_t=\alpha_{t+1}-\eta(\alpha-v_{t+1})\le \alpha_{t+1}-\eta(\alpha-1)<\alpha_{\text{low}}+2\eta(1-\alpha)$, because $v_{t+1} \in \{0, 1\}$. Hence $(\alpha_t-\alpha_{\text{low}})/(\eta(1-\alpha))<2$, which together with~\eqref{eqn:r_t} implies $r_t=1$. Thus~\eqref{eqn:FRI_u_cons_1} enforces $x_{t+1}\in\mathcal{S}_1$, and by~\eqref{eqn:rci} the constraint~\eqref{eqn:FRI_u_cons_2} is feasible.}
    Thus, $(x_{t+1}, \alpha_{t+1}) \in \Pi$. \\
    III. First, from I and II $(x_0, \alpha_0) \in \Pi$ leads to the recursive input feasibility. Second, we prove that $\alpha_t \geq \min\{\alpha_0, \alpha_{\text{low}}\}, \forall t \in \N$ by contradiction. Suppose there  exists $t\in\N$ such that $\alpha_t < \min\{\alpha_0, \alpha_{\text{low}}\}$. Then, at some $\tau<t$, $\alpha_{\tau}$ must satisfy $\min\{\alpha_0, \alpha_{\text{low}}\} \leq \alpha_{\tau} < \min\{\alpha_0, \alpha_{\text{low}}\}+\eta(1-\alpha)$ and \change{$v_{\tau+1}=1$}. However, for such a $\alpha_{\tau}$, \eqref{eqn:FRI_u_cons_2} would be active at time $\tau$, making \change{$v_{\tau+1}=1$} impossible. 
    Finally, the truth that $\alpha_t$ is lower bound by the finite $\min\{\alpha_0, \alpha_{\text{low}}\}$ sufficiently guarantees \textbf{C1}.  The additional $\alpha_0 \leq \alpha_{\text{low}}$ directly leads to \textbf{C2}. \change{Then, \eqref{eqn:cons_vio_asy} and~\eqref{eqn:cons_vio_rob} hold respectively based on Lemma~\ref{lem:bound}}.    
\end{pf}


\change{Intuitively, when the average violation rate becomes too large, the controller leverages the largest disturbance set $\mathbb{W}$ and enforces $x_{t+1}\in\mathcal{X}$ for all $w_t\in\mathbb{W}$.}
The parameter $n_s$ is user-defined.
Due to the nested property that $\mathcal{S}_{k} \subseteq \mathcal{S}_{k+1}$~\citep{korda2012stochastic}, a larger $n_s$ results in a larger maximal set $\mathcal{S}_{k}$. Thus, a larger $n_s$ combined with a smaller $\alpha_{\text{low}}$ expands the feasible region for $\Pi$ due to~\eqref{eqn:r_t}. While Theorem~\ref{thm:FRI} holds for any choices of $n_s$, an inappropriate selection may lead to bad closed-loop control costs. For instance, if $n_s=1$ is used and $\mathcal{S}_{1}$ is relatively small, it may yield very conservative results. 

\begin{remark}
The guarantees in Theorem~\ref{thm:FRI} rely on bounded disturbances $w_t\in\mathbb{W}$ and the resulting RCI sets. The process $w_t$ need not be i.i.d. It may capture exogenous disturbances and bounded model mismatch, as in robust MPC~\citep{marruedo2002input}.
\change{Beyond the nominal linear setting, the framework applies to classes of systems for which RCI sets are computationally obtainable, such as linear systems with parametric uncertainty~\citep{grieder2003robust,liu2019full} and certain nonlinear systems with difference-of-convex dynamics~\citep{fiacchini2010computation}.} Furthermore, the bounded-disturbance assumption can be relaxed by replacing the worst-case RCI set with a backup-policy mechanism, as discussed in~\citep{shi2026disturbance}.
\end{remark}

\change{By Theorem~\ref{thm:FRI}, with a suitable choice of $\alpha_0$, DAD-MPC with $\pi(\cdot)$ defined in~\eqref{eqn:dadmpc_policy} guarantees either~\eqref{eqn:cons_vio_asy} or~\eqref{eqn:cons_vio_rob}. Many components remain design choices, including elements of $\pi(\cdot)$ (horizon, cost, policy structure satisfying~\eqref{eqn:FRI_u_cons}) and the disturbance-set estimator $W(\bar{\alpha}_t)$. The next section provides practical design guidance for $\pi(\cdot)$ and $W(\bar{\alpha}_t)$.}

\subsection{Practical design guidance for DAD-MPC} \label{sect:DAD-MPC}

\change{The components $\pi(\cdot)$ and $W(\bar{\alpha}_t)$ in DAD-MPC can be designed using the following guidance:}
\begin{itemize}[nolistsep,leftmargin=1.5em]
\item Design guidance (\textbf{DG}):
     \change{Make the estimated disturbance set $W(\bar{\alpha}_t)$ and the state constraint tightening in $\pi(\cdot)$ monotonically non-increasing in $\bar{\alpha}_t$.} 
\end{itemize}
Consider a scenario where the average violation over a time period from step $t_1$ to step $t_2$ exceeds the target $\alpha$, i.e. $\frac{\sum_{i=t_1}^{t_2} v_i}{t_2-t_1+1} > \alpha$.
From the update equation~\eqref{eqn:dad_alpha}, we have:
\begin{align*}
    \alpha_{t_2}-\alpha_{t_1-1} & = \eta\sum_{i=t_1}^{t_2} (\alpha - v_i) \\
    & = \eta(t_2-t_1+1)(\alpha - \frac{\sum_{i=t_1}^{t_2} v_i}{t_2-t_1+1}) <0.
\end{align*} 
\change{Hence $\bar{\alpha}_{t_2}\le \bar{\alpha}_{t_1-1}$ (clipping is monotone). If \textbf{DG} holds, a decrease of $\bar{\alpha}_t$ enlarges $W(\bar{\alpha}_t)$ and tightens the state constraints in $\pi(\cdot)$, which in turn reduces violations.
\change{In fact, we have $ \alpha_t = \alpha_0 + \eta(\alpha t - \sum_{i=1}^{t} v_i)$ similarly, which formulates a \textbf{Proportional controller} tracking the ramp signal $\alpha t$}. 
Thus \textbf{DG} induces a violation-feedback loop that regulates the time-averaged violations around the target $\alpha$.}

\change{Since Theorem~\ref{thm:FRI} requires no additional properties of $W(\cdot)$, $W(\cdot)$ may be inexact, enabling data-driven disturbance quantification. As for $\pi(\cdot)$, robust MPC is natural because larger disturbance sets induce tighter constraints.}
This section gives an example that combines conformal prediction~\citep{cp_vovk2005} with disturbance-affine MPC~\citep{goulart2006optimization}. 

\subsubsection{Conformal prediction} \label{sect:CP_for_W}

We next apply split conformal prediction (SCP)~\citep{cp_vovk2005,cp_intro_angelopoulos2021}, a distribution-free and data-driven approach to construct the disturbance estimator.

The quantification of the $j^{th}$-dimension of disturbance $w_t$ using SCP is summarized in Algorithm~\ref{alg:cp_w}. 
In step 1), calibration data are collected by controlling the system~\eqref{eqn:LTI_sys} using a chosen controller $\pi^b$. This controller is user-defined, such as a rule-based controller, nominal MPC or even random signals.
In steps 2)–3), SCP produces a bound for the scalar component $w_t(j)$ using absolute residuals. \change{Finally, $W(\delta)$ is built by a box intersected with $\mathbb{W}$. Note that $\mathcal{C}_{j}(\delta)$ in~\eqref{eqn_w_bound_1} becomes unbounded as $\delta$ approaches $0$, while $W(\delta)$ remains bounded due to the intersection $\mathcal{C}^{w}(\delta)\cap\mathbb{W}$ in~\eqref{eqn:cp_w_bound}.}
SCP provides a probabilistic guarantee under specific conditions, but they are not required by DAD-MPC. \change{For a gentle introduction, see~\citet{cp_intro_angelopoulos2021}}

\begin{algorithm} 
  \caption{Disturbance Quantification by SCP} \label{alg:cp_w}
{
\textbf{Input:} A \change{user-defined} controller $\pi^b$ for data collection\\
\textbf{Output:} Function $\mathcal{C}_{j}(\delta)$ for $(1-\delta)$-confidence disturbance set for $w_t(j), j \in \N_{1}^{n_x}$, and the corresponding disturbance-set estimator $W(\bar{\alpha}_t)$
\begin{itemize} [nolistsep,leftmargin=1.5em]
    \item[1)] Apply $n_{cal}$-step inputs using $\pi^b$ to the system~\eqref{eqn:LTI_sys} starting from some time $t_{c}$.
    Collect the \change{historical} calibration data:
    {\setlength{\abovedisplayskip}{1.6pt}\setlength{\belowdisplayskip}{1.6pt}
    \begin{align*}
        \mathbf{D}_{cal} = \left\{(\begin{bmatrix} x_t \\ u_t \end{bmatrix}, x_{t+1}), \; t \in \N_{t_{c}}^{t_{c}+n_{cal}-1}  \right\}.
    \end{align*} 
    }    
    \item[2)] Compute the residuals for the $j$-th dimension of state, i.e. $x_{t}(j)$, \change{for $j \in \N_{1}^{n_x}$:}
    \begin{align*}
        R_{j,t} = &\lvert x_{t+1}(j) - A(j,:)x_t - B(j,:)u_t\rvert \, , \\ & \text{for} \; t \in \N_{t_{c}}^{t_{c}+n_{cal}-1}.
    \end{align*} 
    \item[3)] \change{Construct function $\mathcal{C}_{j}(\delta)$ for $w_t(j), \delta \in [0,1], j \in \N_{1}^{n_x}$:}
    \begin{equation} \label{eqn_w_bound_1}
        \begin{aligned}
        \mathcal{C}_{j}(\delta) = 
        [-q_{j}(\delta), +q_{j}(\delta)],
        \end{aligned}
    \end{equation}   
    where $q_{j}(\delta) = \lceil (n_{cal}+1)(1-\delta)\rceil$-th smallest value in $\left\{R_{j,t_{c}}, \dots, R_{j, t_{c}+n_{cal}-1}, +\infty \right\}$. 
   \item[4)] \change{Construct $W(\bar{\alpha}_t)$ as:}
\begin{equation} \label{eqn:cp_w_bound}
\begin{aligned}
W(\bar{\alpha}_t) = \begin{cases}
        \mathcal{C}^{w}(\bar{\alpha}_t) \cap \mathbb{W} & \text{if } \bar{\alpha}_t \in [0,1), \\
        \{\mathbf{0}\} & \text{if } \bar{\alpha}_t = 1.     
    \end{cases} 
\end{aligned}    
\end{equation}
where the disturbance set $\mathbb{W}$ is assumed to be known by Assumption~\ref{assump:w_bound} and $\mathcal{C}^{w}(\delta)$ is defined as:
\begin{align*}
\mathcal{C}^{w}(\delta) = \left\{w \middle| 
\begin{aligned}
    w_t(j) & \in \mathcal{C}_{j}\left(\delta\right), \\ &\text{for} \;  j \in \N_{1}^{n_x} 
\end{aligned}\right\}.  
\end{align*} 
\end{itemize}
}
\end{algorithm}



\begin{remark}
\change{Because DAD-MPC does not rely on probabilistic coverage of $W(\cdot)$, any data-driven method that returns a bounded set $W(\cdot)$ with a size parameter can be used. For example, \citet{yeh2024end} learns parametric convex uncertainty sets (e.g., ellipsoids or neural-network-based convex sets) that can better capture correlation. We adopt SCP for simplicity and low computational cost. If there is no calibration data, a practical alternative is the scaled estimator $W(\delta)=\kappa(\delta)\,\mathbb{W}$ with $\kappa(\delta)\in[0,1]$ monotone in $\delta$ (assuming $0\in\mathbb{W}$).}
\end{remark}

\subsubsection{Robust MPC}
Consider a standard affine disturbance feedback policy~\citep{goulart2006optimization}: 
\begin{equation} \label{eqn:robmpc_u_w_feedback}
\begin{aligned} 
   \mathbf{u} &= K\mathbf{w} + \mathbf{\bar{u}}  \\
   &= \begin{bmatrix}
       \mathbf{0} & \mathbf{0} & \dots & \mathbf{0}\\
       K_{1,1} & \mathbf{0} & \dots & \mathbf{0} \\
       \vdots & \ddots & \ddots & \vdots \\
       K_{N-1,1} & \dots & K_{N-1,N-1} & \mathbf{0}
   \end{bmatrix}\mathbf{w} + 
   \begin{bmatrix}
       \bar{u}_1 \\
       \vdots \\
       \bar{u}_N 
   \end{bmatrix},
\end{aligned}
\end{equation}
where $\mathbf{u} := [u_{0|t}^{\top},\dots,u_{N-1|t}^{\top}]^{\top}$ and $\mathbf{w} := [w_{0|t}^{\top},\dots,w_{N-1|t}^{\top}]^{\top}$ denote the $N$-step future input sequence and disturbance, respectively. 
The example $\pi(\cdot)$ is formulated by:
\begin{subequations}\label{eqn:robmpc1}
\begin{align} 
    \min_{\mathbf{u}, K, \mathbf{\bar{u}}} & \sum_{i=0}^{N-1} I(x_{i|t}, u_{i|t}) + \lambda||\sigma_{i|t}||_2^2 \nonumber \\
    \text{s.t.} \; & x_{0|t}=x_t, \nonumber\\
    & x_{i+1|t} =  Ax_{i|t} + Bu_{i|t} + w_{i|t}  \nonumber \\
    & \mathbf{u} = K\mathbf{w} + \mathbf{\bar{u}} \; \text{structured as in}~\eqref{eqn:robmpc_u_w_feedback} \nonumber \\
    & F_x x_i \leq f_x + \sigma_{i|t}, u_i \in \mathcal{U} \label{eqn:robmpc1_soft} \\
    & u_{0|t} \in \mathbb{U}(x_t, \alpha_t) \label{eqn:robmpc1_fri} \\
    &  \forall w_{i|t} \in W(\bar{\alpha}_t) , \forall i \in \N_0^{N-1}. \nonumber
\end{align}
\end{subequations}
Here, the cost function $I(x_{i|t}, u_{i|t})$ is a task-dependent stage cost. The slack variables $[\sigma_{1|t}^{\top},\dots,\sigma_{N|t}^{\top}]^{\top}$ ensure the feasibility of~\eqref{eqn:robmpc1_soft}, penalized by the cost $\lambda||\sigma_{i|t}||_2^2$.
\change{Constraint~\eqref{eqn:robmpc1_fri} embeds the FRI condition, so~\eqref{eqn:robmpc1} belongs to the policy in~\eqref{eqn:dadmpc_policy}.}

\change{The robust mpc~\eqref{eqn:robmpc1} with the SCP-based $W(\bar{\alpha}_t)$ is convex when $I(\cdot)$ is convex.} As $\bar{\alpha}_t$ decreases from $1$ to $0$, $W(\bar{\alpha}_t)$ in~\eqref{eqn:cp_w_bound} expands from $\{\mathbf{0}\}$ by the construction in~\eqref{eqn_w_bound_1}, effectively shifting~\eqref{eqn:robmpc1} from nominal MPC toward a more conservative tightening. Thus the user-defined step size $\eta$ in DAD-MPC strongly affects performance; if $\alpha\in(0,1)$ and $\eta$ is too large, the policy may oscillate between nominal and robust MPC behavior.


\change{DAD-MPC adapts the disturbance set without precise distributional knowledge and leverages the FRI mechanism to prevent violations when the running average violation rate becomes too large. Prior work~\citep{korda2012stochastic,oldewurtel2013adaptively} also uses a similar adaptation approach. However,~\citet{korda2012stochastic} requires known disturbance distributions to set chance constraints, and~\citet{oldewurtel2013adaptively} assumes one-step controllability to guarantee bounds. Our scheme avoids these assumptions. A numerical comparison is as follows.}

\begin{figure*}[!ht]
    \centering
    \includegraphics[width=1.0\linewidth]{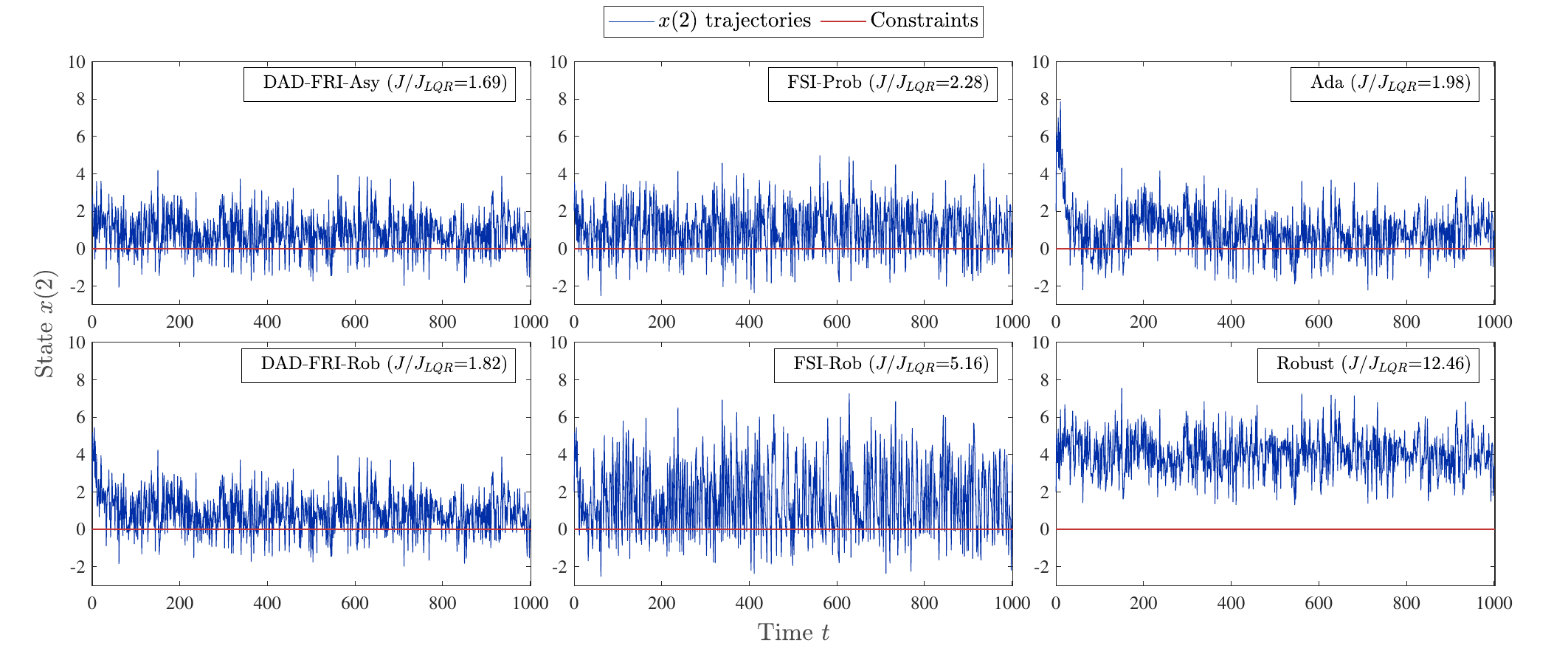}
    \vspace{-12pt} 
    \caption{\footnotesize Comparison of $x(2)$ trajectories by different controllers}
    \label{fig:compare_x2}
\end{figure*}

\begin{figure*}[!ht]
    \centering
    \includegraphics[width=1.0\linewidth]{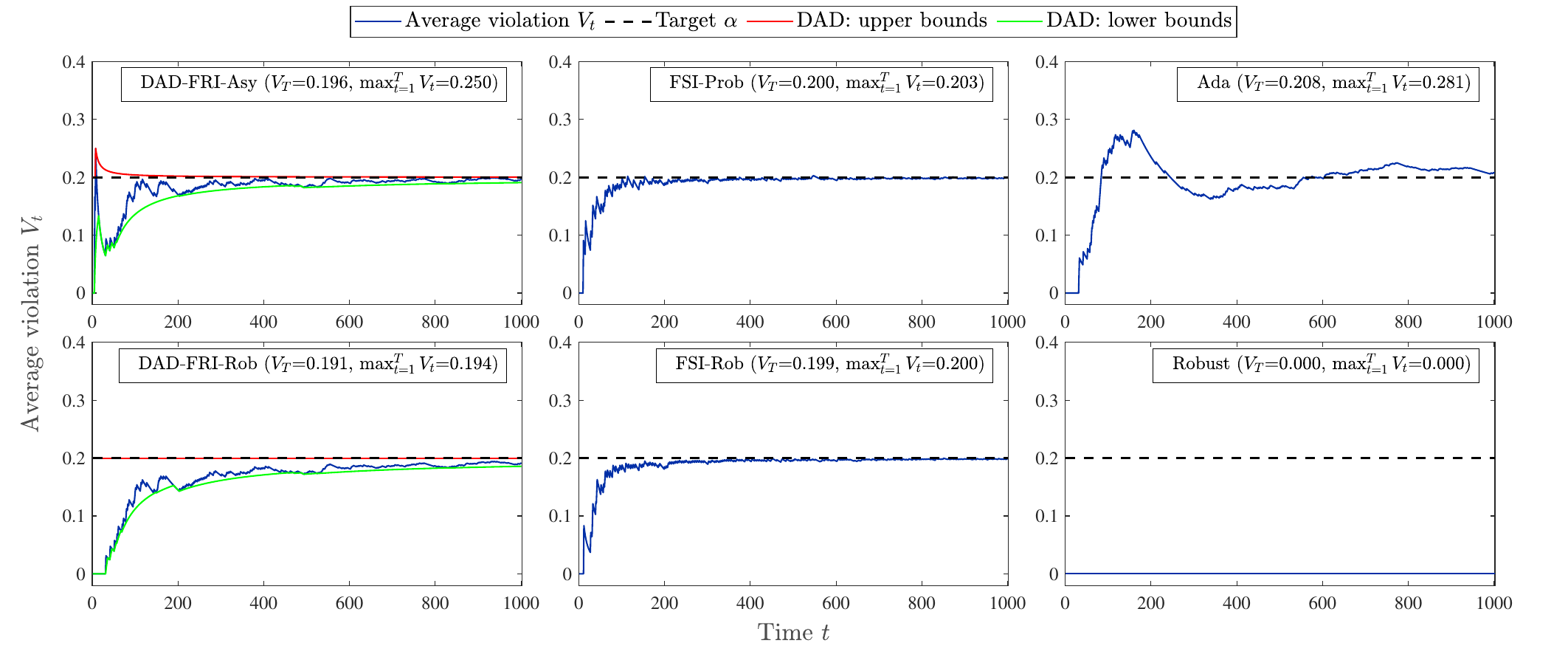}
    \vspace{-12pt} 
    \caption{\footnotesize Comparison of $V_t$ trajectories by different controllers}
    \label{fig:compare_Vt}
\end{figure*}

\section{Simulation validation} 

This section validates the DAD-MPC through simulation. We considered the same setup as in~\citet{korda2012stochastic} and compare the state-of-the-art controllers in~\citet{korda2012stochastic} and~\citet{oldewurtel2013adaptively}. Consider the system~\eqref{eqn:LTI_sys} with the dynamic matrices: $A=\begin{bmatrix}
    1&0\\1&1
\end{bmatrix}, B=\begin{bmatrix}
    1\\0.7
\end{bmatrix}$. The disturbance $w_t$ followed an i.i.d.\ normal distribution $\mathcal{N}(\mathbf{0}, I)$  truncated at 3, i.e. $\lVert w_t \rVert_{\infty}\leq 3$. The inputs and states were constrained by box constraints: $|u| \leq 12, |x(1)|\leq 7, x(2)\geq0$ and $x(2)\leq12$. At time $t$, the closed-loop cost $J$ was calculated using the stage cost, $x^{\top}Qx+u^{\top}Ru$, where $Q=diag(0,1)$ and $R=0.1$. This simulation adopted this cost setup and a horizon length of $N=8$ for MPC-based methods.

For DAD-MPC, \eqref{eqn:robmpc1} was used as $\pi(\cdot)$ with the nominal stage cost. We denote the setup for the asymptotic average violation bound~\eqref{eqn:cons_vio_asy} as DAD-FRI-Asy and the setup for the robust one~\eqref{eqn:cons_vio_rob} as DAD-FRI-Rob. 
We chose $n_s=6, \alpha_{\text{low}}=0$, and chose $\alpha_0=\alpha$ for DAD-FRI-Asy and $\alpha_0=0$ for DAD-FRI-Rob, as indicated in Theorem~\ref{thm:FRI}. We collected 1000-step calibration data by controlling the system~\eqref{eqn:LTI_sys} with an infinite-horizon Linear Quadratic Regulator (LQR). The disturbance set $W(\bar{\alpha}_t)$ was then built as~\eqref{eqn:cp_w_bound}. 
Other setups were settled based on specific control tasks.

For all methods, we ran $T=1000$ steps using the same disturbance realization for fair comparison. We denote $V_t:=\frac{\sum_{i=1}^{t} v_i}{t}$. In the experiments, we report $V_T$ and $\max_{t=1}^{T} V_t$ to observe how the controller behaves relative to the violation bounds in~\eqref{eqn:cons_vio_asy}, \eqref{eqn:cons_vio_rob}, and~\eqref{eqn:cons_vio_milan}. We computed the relative cumulative closed-loop costs as $J/J_{LQR}$ for a clear comparison, where $J_{LQR}$ is the cost from the LQR.

\subsection{A comparison study} \label{sect:sim_A}

We compared DAD-FRI-Asy and DAD-FRI-Rob against other control methods: first-step stochastic (FSI-Prob) and robust invariance methods (FRI-Rob) respectively for the average probability constraint of the violation~\eqref{eqn:cons_vio_milan} and the robust average violation bound~\eqref{eqn:cons_vio_rob} from~\citet{korda2012stochastic}, the adaptively constrained MPC approach (Ada)~\citep{oldewurtel2013adaptively}, the standard affine disturbance feedback robust MPC (Robust). Most methods require the disturbance support, i.e., $\mathbb{W} = \{w\in\R^2 | \lVert w \rVert_{\infty}\leq 3 \}$.  Besides, FSI-Prob requires the correct confidence regions $W(\bar{\alpha}_t)$, formulated by scaled symmetric boxes around the origin according to $w_t$'s distribution~\citep{korda2012stochastic}. 

First, we compared the two DAD-MPC methods with the other controllers for $\alpha=0.2$. The DAD-MPC methods used $\eta=0.1$. Trajectories of the state $x(2)$ and $V_t$ are presented in Fig.~\ref{fig:compare_x2} and~\ref{fig:compare_Vt}.
The two DAD-MPC methods achieved the best control costs with low variance in $x(2)$. Meanwhile, the average violation bounds were satisfied and $V_t$ adheres to the upper and lower bounds established in Lemma~\ref{lem:dad_bound}, as illustrated in Fig.~\ref{fig:compare_Vt}.
In comparison, FSI-Prob leveraged the exact distribution knowledge $W(\bar{\alpha}_t)$ to sufficiently ensure the average probability of violation~\eqref{eqn:cons_vio_milan}. However, the sufficiency can lead to conservatism. FSI-Rob did not leverage any distribution information, resulting in the state jumps across $\mathcal{X}_r$ and different $\mathcal{S}_k$ and therefore high variance in $x(2)$. 
Ada exhibited a notable degree of oscillation in violations and states, which was likely due to its use of an open-loop robust MPC and a multiplicative updating rate. 

\begin{figure}[!ht]
    \centering
    \includegraphics[width=0.8\linewidth]{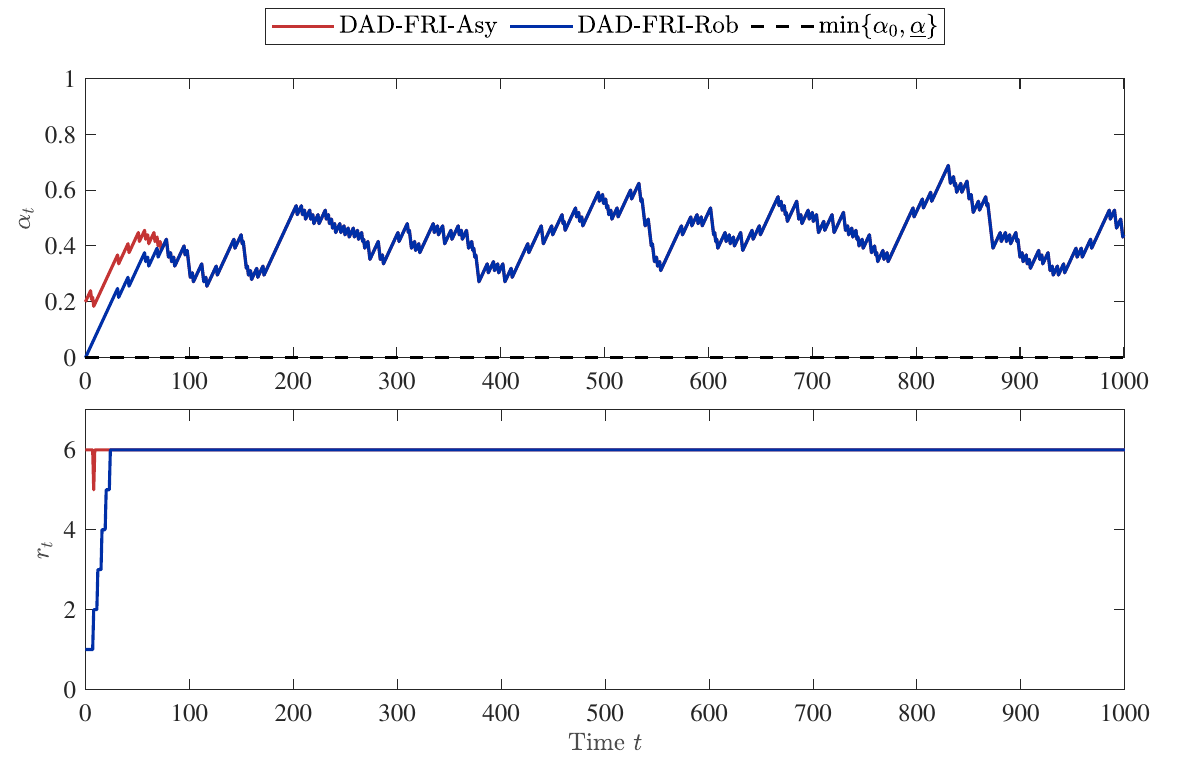}
    \vspace{-12pt} 
    \caption{\footnotesize $\alpha_t$ and $r_t$ Trajectories in two DAD-MPC methods.}
    \label{fig:alpha_t_r_t}
\end{figure}

Furthermore, the trajectories of $\alpha_t$ and $r_t$ of DAD-MPC are presented in Fig.~\ref{fig:alpha_t_r_t}.
It shows that~\eqref{eqn:FRI_u_cons_2} was activated only at the beginning of the DAD-FRI-Rob case when $r_t=1$ because $\alpha_0=\alpha_{\text{low}}=0$.
The largest $\mathcal{S}_6$ was used most of the time, which means the FRI constraint~\eqref{eqn:FRI_u_cons} in~\eqref{eqn:robmpc1} was seldom active.
Instead, both methods
mostly depended on the violation feedback loop based on the $\alpha_t$ adaptation~\eqref{eqn:dad_alpha} and \textbf{DG}. It directly led to the lower-bounded $\alpha_t$ and played an important role in the closed-loop performance.

\begin{table*}[!ht]
\setlength{\tabcolsep}{3pt}
\renewcommand{\arraystretch}{1.7}
\centering 
 \scriptsize 
\caption{\footnotesize Comparison of control methods for different $\alpha$. The criterion includes: (1) the relative cumulative cost $J/J_{LQR}$; (2) the final average violation $V_T$; (3) the maximal average violation in the process $\max_{t=1}^{t} V_t$. Higher $\alpha$ are omitted as the chosen disturbance prevents nominal MPC from violating beyond ~50\%.
}  
\label{tab:compare_alpha}
\begin{tabular}{  *{16}{c}}
 & \multicolumn{3}{c}{$\alpha$=0} &\multicolumn{3}{c}{$\alpha$=0.10} &\multicolumn{3}{c}{$\alpha$=0.20} &\multicolumn{3}{c}{$\alpha$=0.30} &\multicolumn{3}{c}{$\alpha$=0.40} \\
 &$\frac{J}{J_{LQR}}$&$V_T$& $\displaystyle\max_t V_t$ &$\frac{J}{J_{LQR}}$&$V_T$& $\displaystyle\max_t V_t$ &$\frac{J}{J_{LQR}}$&$V_T$& $\displaystyle\max_t V_t$ &$\frac{J}{J_{LQR}}$&$V_T$& $\displaystyle\max_t V_t$ &$\frac{J}{J_{LQR}}$&$V_T$& $\displaystyle\max_t V_t$ \\
  \hline
   LQR &1 &0.490 &0.583 &1 &0.490 &0.583 &1 &0.490 &0.583 &1 &0.490 &0.583 &1 &0.490 &0.583\\
   \hline
DAD-FRI-Asy (Ours) &12.893 &0 &0 
&\underline{\textbf{2.456}}&0.103&0.250
&\underline{\textbf{1.691}}&0.196&0.250 
&\underline{\textbf{1.386}}&0.295&0.301 
&\underline{\textbf{1.157}}&0.395&0.395\\
   DAD-FRI-Rob (Ours) &12.893 &0 &0 
   &2.785 &0.093 &0.095 
   &1.822 &0.191 &0.194 
   &1.451 &0.292 &0.294 
   &1.212 &0.393 &0.393\\
   FSI-Prob &12.893 &0 &0
   &4.082 &0.099 &0.104 
   &2.284 &0.200 &0.203 
   &1.514 &0.300 &0.310 
   &1.168 &0.399 &0.401\\
   FRI-Rob &12.893 &0 &0
   &9.367 &0.099 & 0.100 
   &5.160 &0.199 &0.200 
   &2.993 &0.299 &0.299 
   &1.588 &0.399 &0.399\\
   Ada &23.418 &0 &0 
   &6.855 &0.095 &0.162
   &1.976 &0.208 &0.281 
   &1.680 &0.308 &0.378 
   &1.234 &0.416 &0.427\\
   Robust &\underline{\textbf{12.456}}&0&0 
   &12.456&0&0 &12.456&0&0 &12.456&0&0 &12.456&0&0\\ 
  \hline
\end{tabular}
\end{table*}

Then, we compare performance under different $\alpha$, summarized in Table~\ref{tab:compare_alpha}. For the two DAD-MPC controllers, we used $\eta=1$ if $\alpha=0$ and $\eta=0.5\alpha$ otherwise. The selection is manual, and an optimal $\eta$-tuning study is left as future work. 
The results indicate that all the methods achieved their respective targets on average constraint violations according to the $V_T$ and $\max_{t=1}^{T} V_t$. 
Comparing control costs among these methods, the proposed DAD-FRI-Asy and DAD-FRI-Rob exhibited the best costs in almost all cases for $\alpha\neq0$.  Ada showed worse performance when $\alpha$ was small due to its open-loop robust MPC formulation.

\subsection{Inaccurate model for controller design}

In practice, the actual dynamics~\eqref{eqn:LTI_sys} may not be perfectly identified. We therefore design the controller on the perturbed model:
\begin{equation}
\begin{aligned} \label{eqn:LTI_sys2}
    x_{t+1} = \bar{A}x_t + \bar{B}u_t + \bar{w}_t,
\end{aligned}    
\end{equation}
where the ``total disturbance'' $\bar{w}_t\in\R^{n_x}$ includes model mismatch and exogenous disturbance:
\begin{align*} 
    \bar{w}_t = (A-\bar{A})x_t + (B-\bar{B})u_t + w_t.
\end{align*}
In the simulation, the inaccurate model is generated by $\bar{A}=A+e_A$ and $\bar{B}=B+e_B$, where each entry of $e_A$ and $e_B$ is sampled i.i.d.\ from $\mathcal{U}[-0.1,0.1]$.

Because occasional state-constraint violations are allowed, $x_t$ may leave $\mathcal{X}$ and an a priori analytic bound on $\bar{w}_t$ based on fixed state bounds is not available. For the simulations, we therefore construct a conservative empirical over-approximation of the total-disturbance support via SCP residuals (Algorithm~\ref{alg:cp_w}). Let $R_{j,i} := |x_{i+1}(j) - \bar{A}(j,:)x_i - \bar{B}(j,:)u_i|$. We define
\begin{align*} \label{eqn:cp_w_bound_2}
    \bar{\mathbb{W}} &= \left\{\, w \ \middle|\ 
    \begin{aligned}
    w(j) \in [-\gamma\, q_{\max,j}, \ \gamma\, q_{\max,j}], \ \ j=1,\dots,n_x
    \end{aligned} \right\}, \\
    & \text{where } \ q_{\max,j} := \max_{i=1,\dots,n_{cal}} R_{j,i} \ \text{ and } \ \gamma=1.5 \nonumber
\end{align*}
This yields a bounded box used for controller design. We verified offline that $\bar{\mathbb{W}}$ contained all realized $\bar{w}_t$ in every Monte Carlo trial.

\begin{figure}[!ht]
    \centering
    \includegraphics[width=1.0\linewidth]{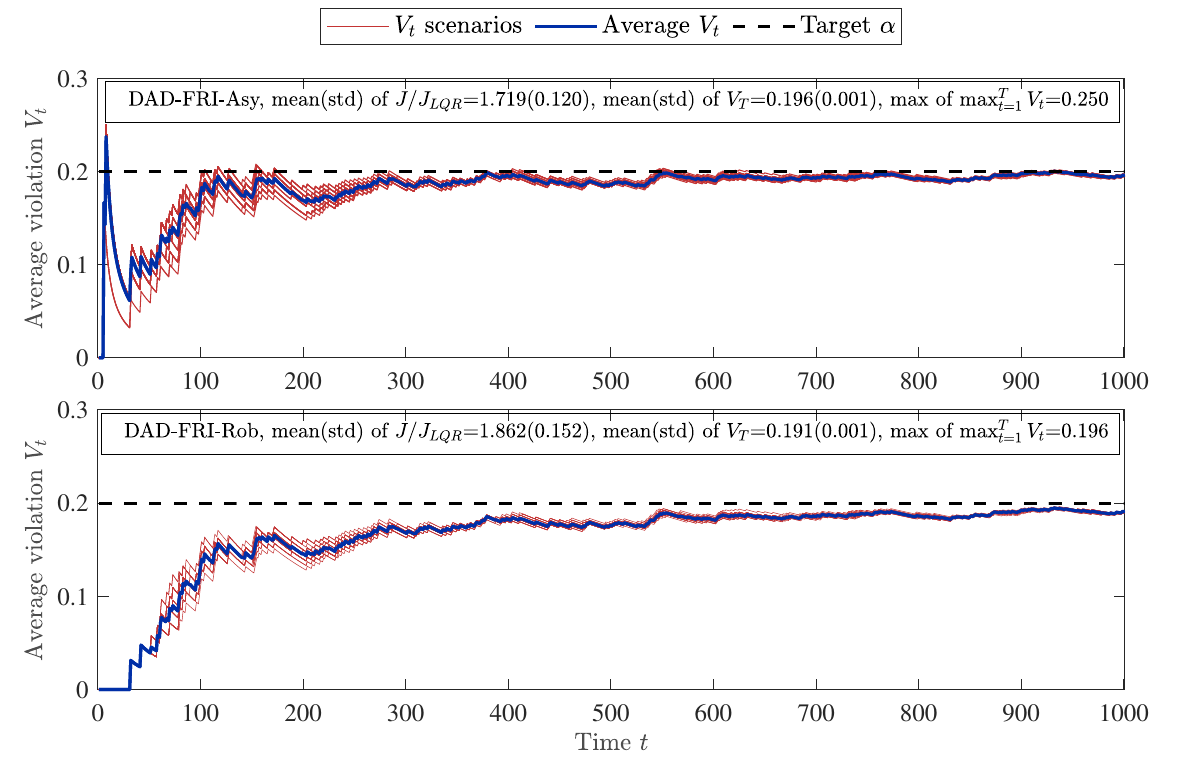}
    \caption{\footnotesize Comparison of violation $V_t$ trajectories between two DAD-MPC methods when considering model errors. Red lines: results from different Monte-Carlo runs; Blue lines: average $V_t$ over all Monte-Carlo runs.}
    \label{fig:model_error}
\end{figure}

We performed 30 Monte Carlo trials with independently sampled $(e_A,e_B)$. Fig.~\ref{fig:model_error} shows the trajectories of $V_t$ for the two DAD-MPC variants: red curves correspond to individual trials and blue curves to the trial averages. Statistics of the closed-loop cost $J/J_{\mathrm{LQR}}$ are also reported.

Across all trials, the violation bounds~\eqref{eqn:cons_vio_asy} and~\eqref{eqn:cons_vio_rob} were satisfied by DAD-FRI-Asy and DAD-FRI-Rob, respectively. Although many sampled inaccurate \emph{open-loop} models are unstable, the DAD-MPC controllers maintain costs $J/J_{\mathrm{LQR}}$ comparable to those obtained with the exact model in Section~\ref{sect:sim_A}. This indicates that DAD-MPC tolerates non-i.i.d., model-dependent disturbance and mitigates conservatism even when an over-estimated disturbance set $\bar{\mathbb{W}}$ is used.



\subsection{Sensitivity analysis}

\begin{figure}[!ht]
    \centering
    \includegraphics[width=0.9\linewidth]{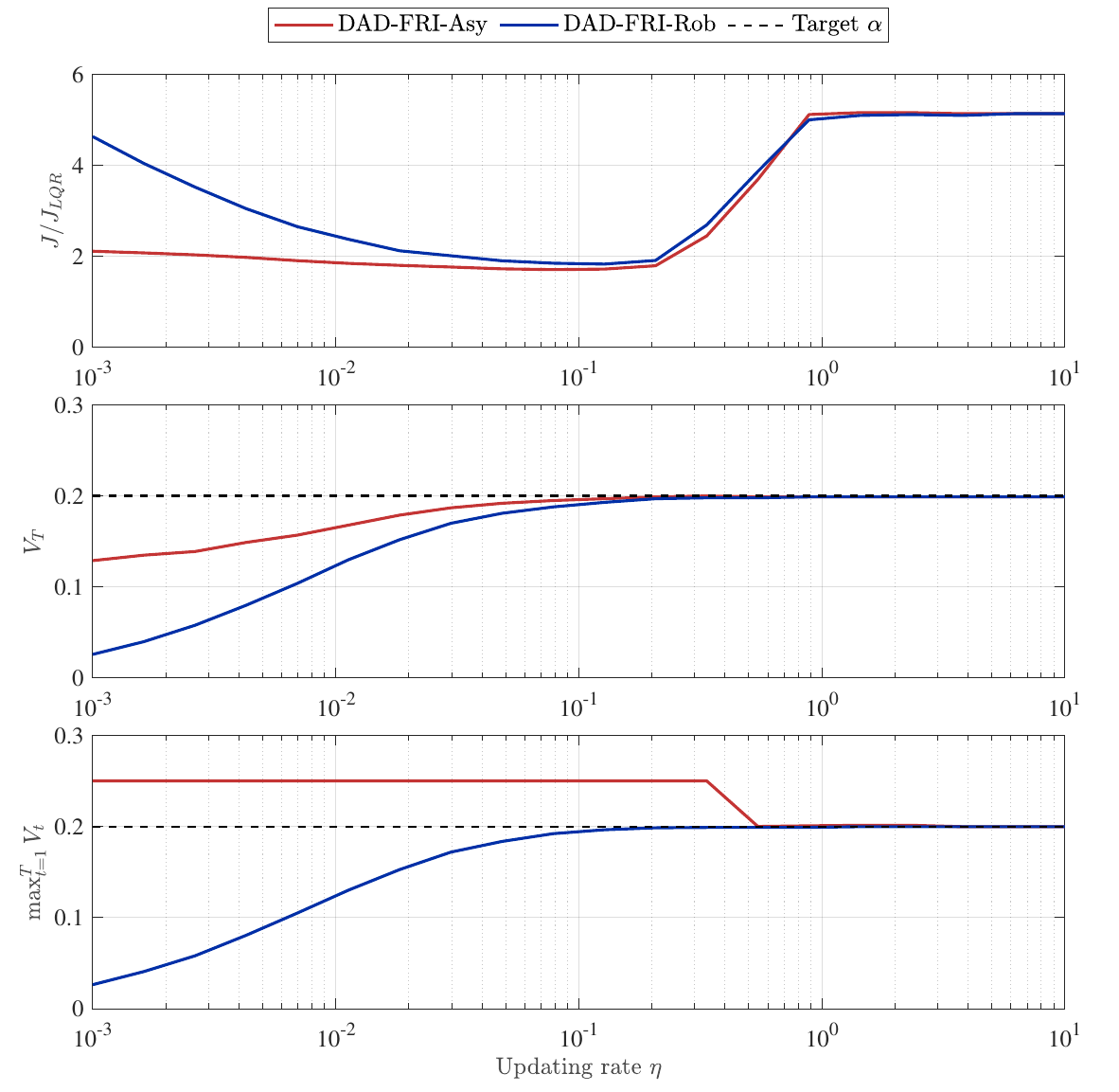}
    \vspace{-12pt}
    \caption{\footnotesize Sensitivity analysis of $\eta$ for two DAD-MPC methods. Top: cumulative control costs Bottom: average violations at the final time step}
    \label{fig:dad_sensi}
\end{figure}

A sensitivity analysis of the updating rate $\eta$ in the DAD-MPC used in Section~\ref{sect:sim_A} is presented in Fig.~\ref{fig:dad_sensi}, highlighting its significant impact on control performance.
The metrics $V_T$ and $\max_{t=1}^{T} V_t$ confirm that the average violation bounds are satisfied in both methods, regardless of the chosen $\eta$.
If $\eta$ is too small, $W(\bar{\alpha}_t)$ adjusts slowly, potentially making $V_T$ too small, as indicated in Lemma~\ref{lem:dad_bound}: \eqref{eqn:thm_bound}, and leading to conservative $J/J_{LQR}$ values. Conversely, if $\eta$ is too large, $J/J_{LQR}$ remains conservative because the constraint~\eqref{eqn:FRI_u_cons_1} becomes dominant, causing state jumps among $\mathcal{S}_k$.
When a suitable updating rate is chosen,  DAD-MPCs effectively adjust the disturbance bounds, achieving optimal $J/J_{LQR}$ values.

\section{Conclusion}

This work proposes a DAD-MPC framework, which adapts the disturbance set according to the current constraint violation. 
By combining the FRI method, the DAD-MPC is recursively feasible and guarantees asymptotic or robust bounds on the average constraint violation.
Notably, the DAD-MPC controller does not require the exact knowledge of the disturbance distribution and does not need an i.i.d.\ disturbance assumption, showing the potential extension to MPC with data-driven models. 


\section*{DECLARATION OF GENERATIVE AI AND AI-ASSISTED TECHNOLOGIES IN THE WRITING PROCESS}
During the preparation of this work the authors used ChatGPT in order to improve the readability and language and to check spelling and grammar. After using this tool/service, the authors reviewed and edited the content as needed and take full responsibility for the content of the publication.

\bibliography{ref}             
\end{document}